\documentclass[PRL,reprint,amsmath,amssymb,superscriptaddress,longbibliography]{revtex4-1}

\usepackage{graphicx}
\usepackage{dcolumn}
\usepackage{bm}
\usepackage{multirow}
\usepackage{enumitem}

\begin{document}

\title{Chiral anomaly induced negative magnetoresistance in topological Weyl semimetal NbAs}

\author{Xiaojun Yang}
 \affiliation{Department of Physics and State Key Laboratory of Silicon Materials, Zhejiang University, Hangzhou 310027, China}
\author{Yupeng Li}
 \affiliation{Department of Physics and State Key Laboratory of Silicon Materials, Zhejiang University, Hangzhou 310027, China}
\author{Zhen Wang}
 \affiliation{Department of Physics and State Key Laboratory of Silicon Materials, Zhejiang University, Hangzhou 310027, China}
\author{Yi Zheng}
\email{phyzhengyi@zju.edu.cn}
 \affiliation{Department of Physics and State Key Laboratory of Silicon Materials, Zhejiang University, Hangzhou 310027, China}
 \affiliation{Collaborative Innovation Centre of Advanced Microstructures, Nanjing 210093, P. R. China}
 \affiliation{Zhejiang California International NanoSystems Institute, Zhejiang University, Hangzhou 310027, P. R. China}
\author{Zhu-an Xu}
 \email{zhuan@zju.edu.cn}
 \affiliation{Department of Physics and State Key Laboratory of Silicon Materials, Zhejiang University, Hangzhou 310027, China}
 \affiliation{Collaborative Innovation Centre of Advanced Microstructures, Nanjing 210093, P. R. China}
 \affiliation{Zhejiang California International NanoSystems Institute, Zhejiang University, Hangzhou 310027, P. R. China}
\date{\today}

\begin{abstract}
In this paper, we report the intercone transport of Weyl fermions in NbAs with external magnetic field in parallel to electric field, a quantum phenomenon known as the Adler-Bell-Jackiw anomaly. Surprisingly, the resulting negative magnetoresistance (MR) in NbAs shows significant difference from NbP. The observed low-field positive MR dip, which is missing in NbP at low temperatures, indicates that the spin-orbital coupling (SOC) is significantly stronger in NbAs than in the former. The results imply that the contribution of arsenic to SOC in TaAs and NbAs is not negligible. 
\end{abstract}


\maketitle

Since the discovery of two-dimensional (2D) massless Dirac fermions in graphene \cite{Graphe_Berryphase1,Graphe_Berryphase2}, many research efforts have been made in search of topological materials for hosting the other two types of relativistic Weyl fermions and Majorana fermions. Although the observation of Majorana fermions are still under debate, there are recent major breakthroughs in the low-energy quasiparticle realization of chiral Weyl fermions in non-centrosymmetric transition metal monopnictides \cite{WSM_PRX,TaAs_arXiv_NMR,TaAs_arXiv_Jia,TaAs_ARPES1,TaAs_ARPES2,NbAs_ARPES,NbP_arXivWZ} . Unlike the previous proposal of magnetic Weyl semimetals (WSMs) in pyrochlore iridates \cite{pair_PRB}, the spin degeneracy in the TaAs family is lifted by breaking inversion symmetry \cite{WSM_PRX} instead of time reversal symmetry. The non-magnetic, binary compounds of TaAs, TaP, NbAs and NbP are ideal for exploring exotic Weyl fermion-related physics and novel device concepts \cite{chiral_anomaly_PLB,nov_p_nc,nov_p_PRB,nov_p_PRB2,nov_p_PRB3,nov_p_PRX}. Indeed, surface Fermi arcs associated to the topological Weyl node pair with opposite chirality in the bulk have soon been observed in TaAs and NbAs by angle-resolved photoemission spectroscopy (ARPES) \cite{TaAs_ARPES1,TaAs_ARPES2,NbAs_ARPES}, and the chiral-anomaly induced quantum transport \cite{chiral_anomaly_PLB,CME_PRD} has also been demonstrated in TaAs \cite{TaAs_arXiv_NMR} and NbP \cite{NbP_arXivWZ}.

Although all the synthesized TaAs family show chiral Weyl nodes in the bulk, it is not clear at the moment which compound is the best platform for testing various quantum phenomena and device prototypes. Ideally, with the presence of topological Weyl nodes, strong spin-orbital coupling (SOC) is required to fully open an energy bandgap in the bulk and leave the Fermi energy ($E_{F}$) lying within the relativistic Weyl cones. TaAs was naturally regarded as a better choice than NbP, due to the much heavier atomic weight of Ta (5\textit{d}) than Nb (4\textit{d}). However, very recent transport experiments show that NbP has unprecedented Weyl electron mobility of $1\times10^{7}$ cm$^{2}$V$^{-1}$s$^{-1}$ at 1.5 K \cite{NbP_arXivWZ}, which is at least one order of magnitude higher than TaAs \cite{TaAs_arXiv_NMR,TaAs_arXiv_Jia}. NbAs thus becomes the unique choice to understand the difference between TaAs and NbP, since TaP has not be synthesised at ambient pressure. Equally importantly, despite that ARPES results have confirmed Weyl nodes and Fermi arcs in NbAs \cite{NbAs_ARPES}, the quantum signature of negative MR=$[\rho(H)-\rho(0)]/\rho(0)$ associated to chiral Weyl fermions have not been reported yet.

Chiral anomaly induced negative MR in TaAs is characterised by sharp positive MR dip below 0.5 T, which is attributed to SOC induced weak antilocalizaiton (WAL) \cite{WAL_PRL}. Such positive MR dip is completely covered by the negative Adler-Bell-Jackiw anomaly in NbP, and only becomes discernible at the elevated temperatures above 50 K when the systems are not well quantized \cite{NbP_arXivWZ}. Here we report the observation of the competition between chiral anomaly induced negative MR and positive MR dip at low temperatures in NbAs. Despite that Nb is much lighter than Ta, NbAs shows negative MR characteristics more like TaAs than NbP. The results are direct evidence on the non-negligible contribution of As to SOC in TaAs and NbAs.

NbAs crystallizes in a body-centered tetragonal Bravais lattice with the space group of I4$_1$md (109). Our X-ray diffraction (XRD) obtains lattice constants of $a$ = 3.45 {\AA}  and $c$ = 11.68 \AA, consistent with the earlier crystallographic studies\cite{NbAs_crystal1,NbAs_crystal2,NbAs_ARPES}.
Single crystals of NbAs were grown by vapor transport using iodine as the transport agent, as described in Ref. \cite{NbP_arXivWZ,NbP_arXiv}. The chemical compositions of NbAs were verified by energy dispersive x-ray spectroscopy (EDX), showing an atomic
percentage ratio of Nb:As = $49.4:50.6\pm3\%$ without iodine residual in these single crystalsimpurity.
The largest natural surface of the obtained NbAs single crystals was determined to be the (112) plane by single crystal x-ray diffraction, shown in the lower inset of Fig. 1(a), with typical dimension of 1 $\times$ 1 mm$^2$. The quality of the NbAs single crystals was further
checked by the x-ray rocking curve. The full width at half maximum (FWHM) is only 0.03$^{\circ}$ (not showing here), indicating the high quality of the single crystals. The sample was polished to a bar shape, with 1 $\times$ 0.3 mm$^2$ in the (112) plane and a 0.2 mm thickness.
A standard six-probe method was used for both the longitudinal resistivity and transverse Hall
resistance measurements.

\begin{figure}
\includegraphics[width=8cm]{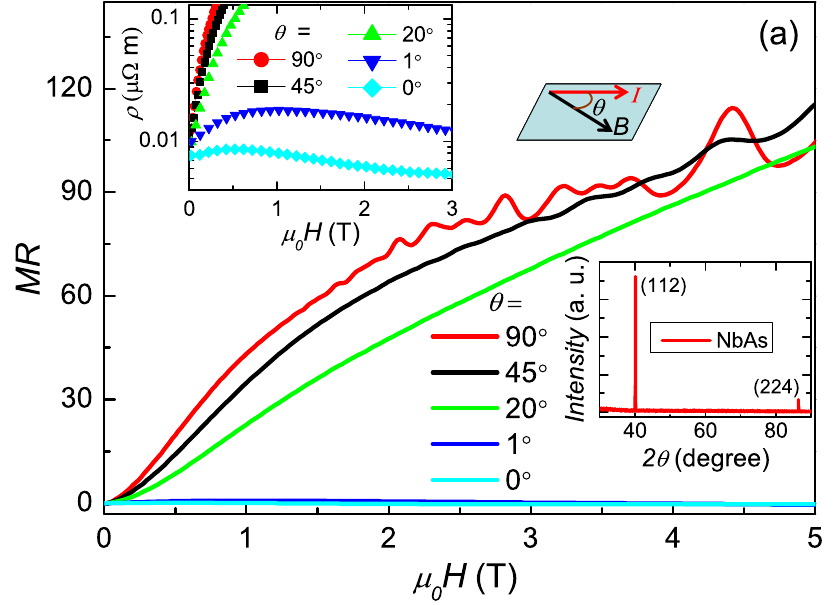}
\includegraphics[width=8cm]{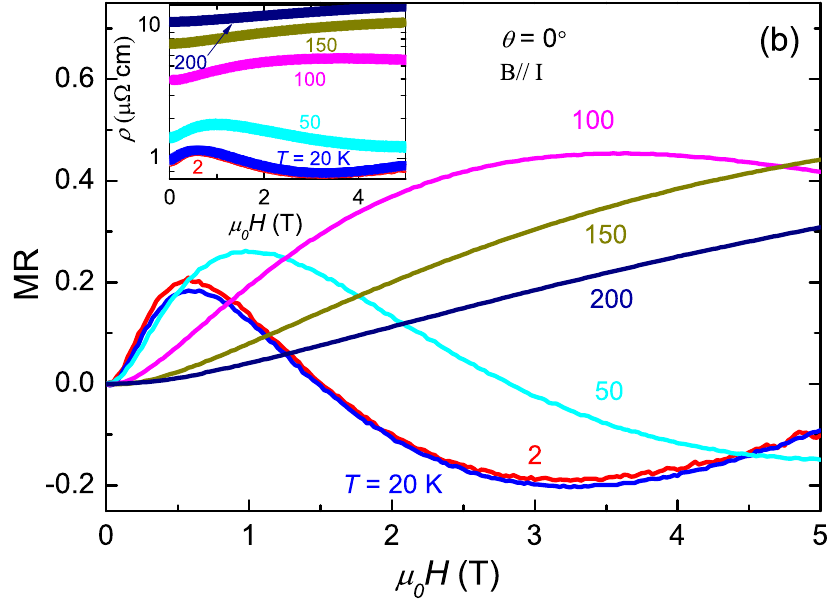}
\caption{\label{Fig.1} (Color online)  (a), Magnetic field dependence of Magnetoresistance
 with magnetic field ($\mu_0H$) from perpendicular ($\theta$ = 90$^\circ$)
 to parallel ($\theta$ = 0$^\circ$) to the electric current ($I$) at $T$ = 2 K. The upper left inset
 displays the original resistivity data plotted on logarithmic scale, emphasizing the
 contrast between extremely large positive MR for magnetic field perpendicular
 to current ($\theta$ = 90$^\circ$) and negative MR for field parallel to
 current ($\theta$ = 0$^\circ$). The upper right  inset depicts the corresponding
 measurement configurations. The lower inset present the single crystal XRD data. (b), Magnetic field dependence
 of MR under various temperatures for magnetic field ($\mu_0H$) parallel
 ($\theta$ = 0$^\circ$) to the electric current ($I$). The inset display the
 original resistivity data.}
\end{figure}

Figure 1(a) displays the angle dependent MR measured at $T$ = 2 K, in which the angles $\theta$ is defined by the applied magnetic field ($\mu_0H$) with respect to the electric current ($I$), as illustrated in the right inset of Fig. 1(a). The angle rotates from $\mu_0H//I$ to $\mu_0H\bot I$, and $\theta$ = 0$^\circ$ corresponds to the former configuration. With $\theta$ = 90$^\circ$ ($\mu_0H\bot I$), positive MR of $\sim$10000\% is
observed, which is strongly relies on $\theta$. When the magnetic field is rotated to be parallel to the electric current ($\theta$ = 0$^\circ$),
we observed the Adler-Bell-Jackiw anomaly associated to Wely fermions in NbAs, manifested as negative MR below 3 T. 
In the left inset of Fig. 1(a), we display the original resistivity data plotted on logarithmic scale, which clearly shows that negative MR is unique for $\theta$ = 0$^\circ$. Since the Adler-Bell-Jackiw anomaly is a quantum effect, we have studied the negative MR at a function of temperatures, by keeping magnetic field parallel to the current($\mu_0H//I$, $\theta = 0^\circ$ ). As shown in Figure 1b, below 20 K, an negative MR change of about -20\% can be
observed under an applied field of 3.2 T. Such negative MR is suppressed by increasing temperature, and ultimately disappeared above 100 K. The T-dependent changes in negative MR is very similar to the results obtained on TaAs and NbP \cite{TaAs_arXiv_NMR,TaAs_arXiv_Jia, NbP_arXivWZ}. At low field (B$<0.5$T), the data show positive MR dip, which may come from the WAL effect stemming from the strong spin-orbit interactions\cite{WAL_PRL}. Similar low-field positive MR has also been reported in TaAs \cite{TaAs_arXiv_NMR,TaAs_arXiv_Jia}, but in stark contrast to NbP in which positive MR dip is completely missing below 75 K \cite{NbP_arXivWZ}. This is quite surprising, since SOC is generally considered to be controlled by the transition metals. It is noteworthy that when magnetic field higher than 3.2 T, the MR becomes positive again. This behavior is very similar to the situation in TaAs\cite{TaAs_arXiv_NMR}. One possible explanation was given by Huang $et al.$. in Ref. \cite{TaAs_arXiv_NMR} as the Coulomb
interaction among the electrons occupying the chiral states, which drives the systems into  a spin-density-wave (SDW)
like state.

\begin{figure}
\includegraphics[width = 8cm]{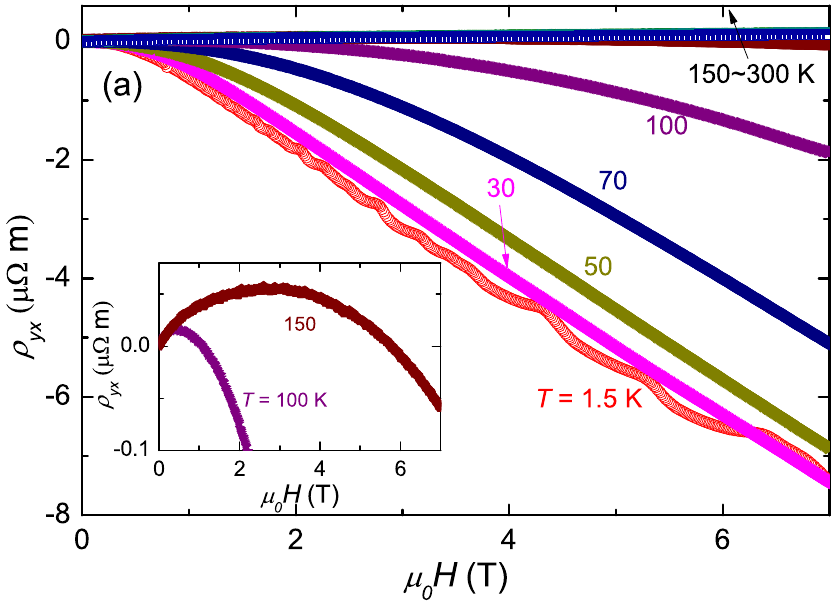}
\includegraphics[width = 8cm]{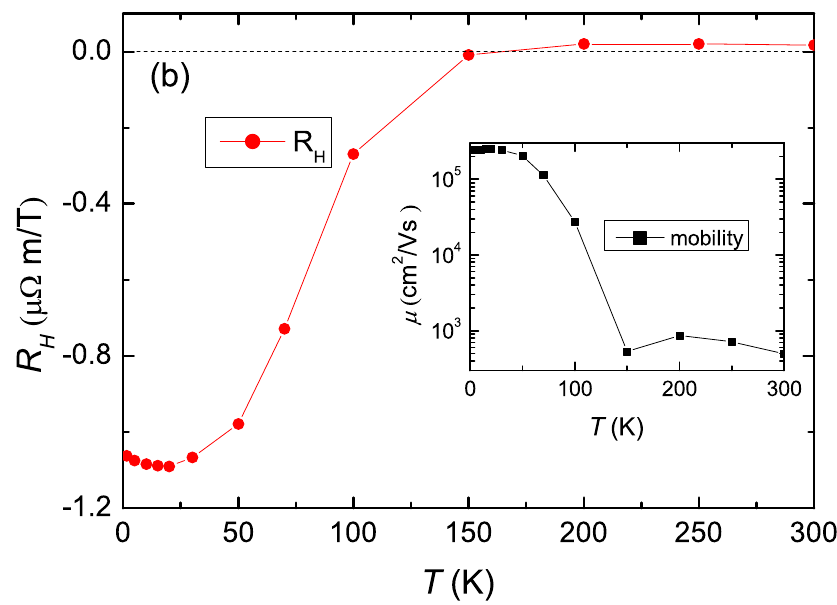}
\caption{\label{Fig.2}(Color online) (a), Hall resistivity measured at various
temperatures from 2 to 300 K. The inset displays enlarged plot of
Hall resistivity curves at $T$ = 100 K and 150 K. (b), Hall coefficient $R_H$ at 7 T
as a function of temperature. The inset displays mobility versus
temperature determined by the Hall coefficient at 7 T and the zero field resistivity
using a single band approximation.}
\end{figure}

Figure 2(a) displays the magnetic field dependence of Hall resistivity $\rho_{yx}$(B//c) measured as a function of temperatures. At low temperature, the negative slope in high magnetic fields indicates that the electrons dominate the charge transport. Insterestingly, in low fields the curve shows a pronounced sign change from positive to negative, as shown in the inset of Figure 2a for 100 K and 150 K. The curvature and sign reversal of the Hall resistivity is typical for the systems with the coexistence of high-mobility electrons carriers with low-mobility holes. At higher temperature, the slope of Hall
resistivity changes to positive, implying the dominant hole conductions. All these results are consistent with multiple hole- and electron-pockets in TaAs and NbP, and agreeing with DFT band structure calculations\cite{NbP_arXiv,TaAs_arXiv_Jia,TaAs_arXiv_NMR,WSM_PRX}
and a previous paper on NbAs\cite{NbAs_jpcm}. As shown in  Fig. 2(b), the material shows negative Hall coefficient, $R_H(T)$ below 150 K, and changes $R_H(T)$ sign for temperature at higher temperatures. For simplicity, we have used the single  band theory to estimate the mobility. The inset of Fig 2(b) displays mobility versus temperature deduced from $R_H(T)$  at 7 T.  Here, NbAs exhibits an ultrahigh mobility of 2.45 $\times$ 10$^5$ cm$^2$/Vs at 1.5 K, consistent with the result in ref. \cite{NbAs_jpcm}.

\begin{figure}
\includegraphics[width=8cm]{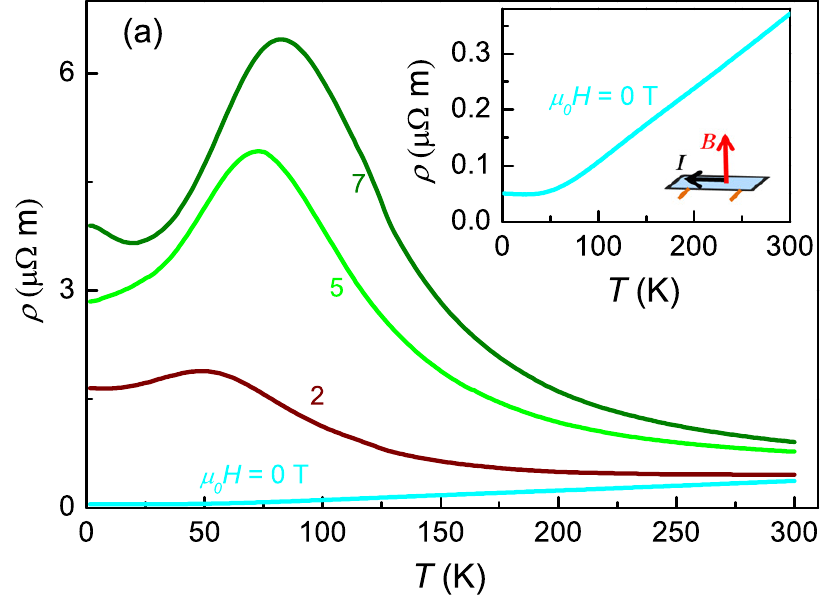}
\includegraphics[width=8cm]{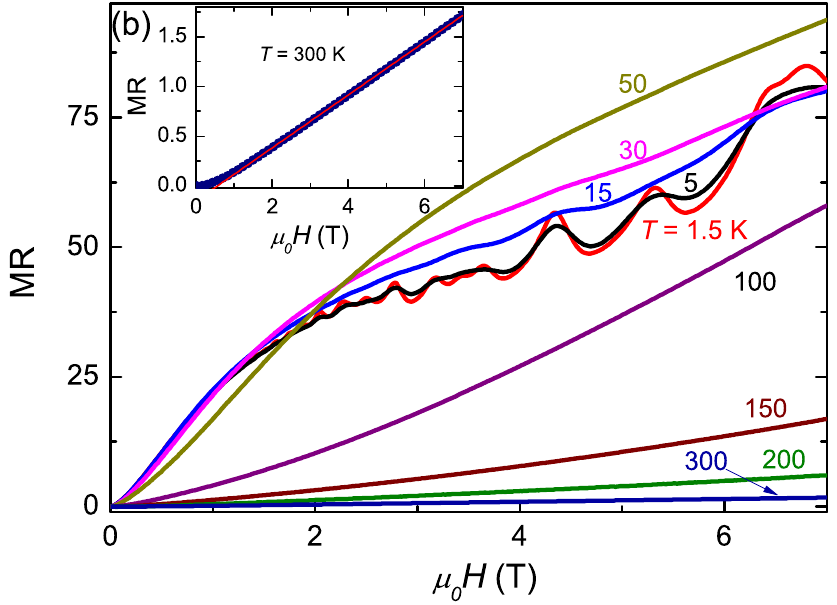}
\caption{\label{Fig.3}(Color online)  (a), The temperature dependence of
resistivity in magnetic field perpendicular to the electrical current.
The inset of (a) gives the measurement configuration, and zooms in on the case of 0 T.
(b) Magnetic field dependence of MR at representative temperatures. The inset
of (b) displays the enlarged plot for $T$ = 300 K.}
\end{figure}

In Fig. 3(a) the temperature dependence of resistivity is plotted. The inset shows the zoom-in of metallic characteristic of NbAs at zero field down to 1.5 K.
The applied magnetic field not only significantly increases the resistivity, but also causes a crossover from metallic to insulator-like behavior, which may be related to the formation of the Landau levels under magnetic field\cite{TaAs_arXiv_NMR}. Noticeably, a resistivity anomaly can be observed under applied field, with its peak position shifting to high temperature when increasing magnetic fields. Similar behavior was observed in TaAs\cite{TaAs_arXiv_NMR}, ZrTe$_5$ and HfTe$_5$\cite{Zr/HfTe5_anom_PRB}. The mechanism may be closely related to the origin of linear MR and detailed study is still under the way. Figure 3(b) displays the field dependence of MR at various temperatures. At 300 K, the MR reaches as high as 170\% at 7 T, as shown in the
inset of Fig. 3(b). Above 1.5 T, the MR of our NbAs single crystal is quite linear, which can be attributed to the theoretical prediction of  linear quantum MR induced by 3D relativistic electronic structure\cite{Cd3As2_theory_LMR}. However, the quantum limit required in Ref.\cite{Cd3As2_theory_LMR} is actually not reached in our sample. Nevertheless, this large room temperature linear MR is quite unusual. At low temperatures, clear Shubnikov de Haas (SdH) oscillations have been detected starting from very weak magnetic field.

\begin{figure}
\includegraphics[width=8cm]{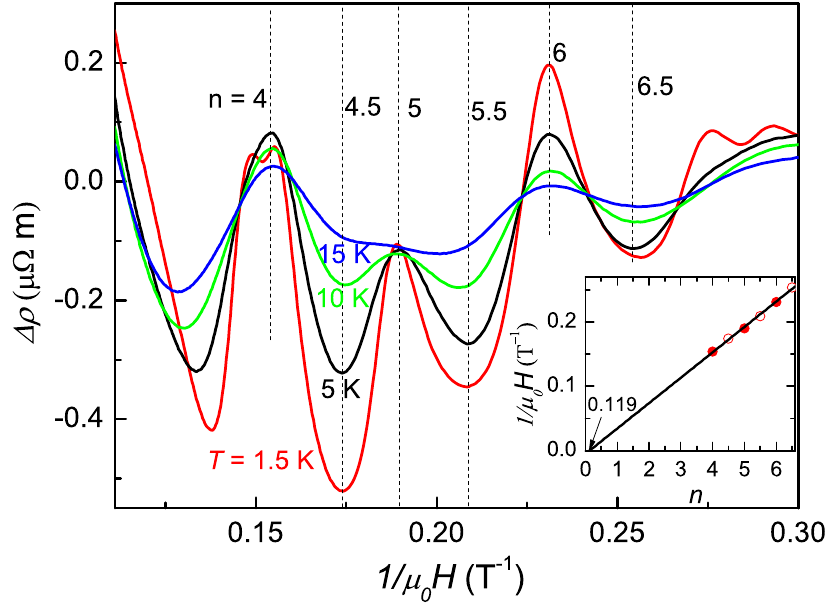}
\caption{\label{Fig.4}(Color online)  The high field
oscillatory component $\vartriangle\rho$ plotted against inverse field 1/$\mu_0H$ at
$T$ = 1.5, 5, 10 and 15 K.  The inset displays SdH fan
diagram plotting the measured1/$B_n$ with the filling factor n, which
is estimated from the $\vartriangle\rho$ versus 1/$B_n$ plot.  }
\end{figure}

Figure 4 shows the oscillatory component of $\Delta\rho$ versus $1/B$ at different temperatures after subtracting polynomial backgrounds. The deduced $\Delta\rho$ versus $1/B$ curves are periodic in $1/B$, corresponding to the successive emptying of the Landau levels when the magnetic field is increased. The peaks marked by Landau level (LL) index n are mainly stemming from the oscillations with the frequency of 25 T.  The cross-sectional area of the Fermi surface (FS) $A_F = 2.38 \times 10^{-3} \text{\AA}^{-2}$ can be obtained according to the Onsager relation $F = (\Phi_0/2\pi^2)/A_F$, where $F$ and $\Phi_0$ are the oscillation frequency and flux quantum, respectively. We assign integer indices to the $\Delta\rho$ peak positions in $1/B$ and half integer indices to the $\Delta\rho$ valley positions. According to the Lifshitz-Onsager quantization rule $A_F(\hbar/eB) = 2\pi(n+\beta-1/2+\delta)$, in which 2$\pi\beta$ is the Berry's phase, and 2$\pi\delta$ is an additional phase shift resulting from the three dimensionality of the Fermi surface \cite{delta_science}, the LL $n$ is linearly dependent on $1/B$.  The fitting of our results in Fig. 4 yields an intercept 0.119, which means a non-trivial $\pi$ Berry's phase ($\beta = 1/2$) and $\delta = 0.119$ (very close to 1/8), which is expected for electron pockets enclosing Weyl nodes.


In summary, we have performed bulk transport measurements on single crystals of 3D Weyl semimetal NbAs, with solid quantum transport evidence on the existence of Weyl fermion pockets. Large MR as high as 10000\% is detected with magnetic field perpendicular to the current. When the external magnetic field is rotated to be parallel with the current, chiral anomaly induced negative MR up to -20\% is observed, providing the first experimental results on quantum transport of Weyl fermions in NbAs. The decent mobility of 2.45 $\times$ 10$^5$ cm$^2$/Vs at 1.5 K in NbAs allow us to analyze the Shubnikov-de Haas oscillations of longitudinal resistance as a function of temperatures, which yields a nontrivial $\pi$ Berry's phase as expected for the relativistic Weyl cone. Unexpectedly, NbAs show much stronger SOC than NbP, as manifested by positive MR dips below 0.5 T, which competes with the chiral anomaly induced negative MR at low temperatures. The results suggest that the contribution of As to  SOC is not negligible in TaAs and NbAs.

This work is supported  by the National Basic Research Program of China (Grant Nos. 2014CB921203 and 2012CB927404), NSF of China
(Contract Nos. 11190023). Y.Zheng acknowledges the start funding support from the 1000 Youth Talent Program.

\bibliography{NbAsV1ZY}

\begin{thebibliography}{25}%
\makeatletter
\providecommand \@ifxundefined [1]{%
 \@ifx{#1\undefined}
}%
\providecommand \@ifnum [1]{%
 \ifnum #1\expandafter \@firstoftwo
 \else \expandafter \@secondoftwo
 \fi
}%
\providecommand \@ifx [1]{%
 \ifx #1\expandafter \@firstoftwo
 \else \expandafter \@secondoftwo
 \fi
}%
\providecommand \natexlab [1]{#1}%
\providecommand \enquote  [1]{``#1''}%
\providecommand \bibnamefont  [1]{#1}%
\providecommand \bibfnamefont [1]{#1}%
\providecommand \citenamefont [1]{#1}%
\providecommand \href@noop [0]{\@secondoftwo}%
\providecommand \href [0]{\begingroup \@sanitize@url \@href}%
\providecommand \@href[1]{\@@startlink{#1}\@@href}%
\providecommand \@@href[1]{\endgroup#1\@@endlink}%
\providecommand \@sanitize@url [0]{\catcode `\\12\catcode `\$12\catcode
  `\&12\catcode `\#12\catcode `\^12\catcode `\_12\catcode `\%12\relax}%
\providecommand \@@startlink[1]{}%
\providecommand \@@endlink[0]{}%
\providecommand \url  [0]{\begingroup\@sanitize@url \@url }%
\providecommand \@url [1]{\endgroup\@href {#1}{\urlprefix }}%
\providecommand \urlprefix  [0]{URL }%
\providecommand \Eprint [0]{\href }%
\providecommand \doibase [0]{http://dx.doi.org/}%
\providecommand \selectlanguage [0]{\@gobble}%
\providecommand \bibinfo  [0]{\@secondoftwo}%
\providecommand \bibfield  [0]{\@secondoftwo}%
\providecommand \translation [1]{[#1]}%
\providecommand \BibitemOpen [0]{}%
\providecommand \bibitemStop [0]{}%
\providecommand \bibitemNoStop [0]{.\EOS\space}%
\providecommand \EOS [0]{\spacefactor3000\relax}%
\providecommand \BibitemShut  [1]{\csname bibitem#1\endcsname}%
\let\auto@bib@innerbib\@empty
\bibitem [{\citenamefont {Novoselov}\ \emph {et~al.}(2005)\citenamefont
  {Novoselov}, \citenamefont {Geim}, \citenamefont {Morozov}, \citenamefont
  {Jiang}, \citenamefont {Katsnelson}, \citenamefont {Grigorieva},
  \citenamefont {Dubonos},\ and\ \citenamefont {Firsov}}]{Graphe_Berryphase1}%
  \BibitemOpen
  \bibfield  {author} {\bibinfo {author} {\bibfnamefont {K.~S.}\ \bibnamefont
  {Novoselov}}, \bibinfo {author} {\bibfnamefont {A.~K.}\ \bibnamefont {Geim}},
  \bibinfo {author} {\bibfnamefont {S.~V.}\ \bibnamefont {Morozov}}, \bibinfo
  {author} {\bibfnamefont {D.}~\bibnamefont {Jiang}}, \bibinfo {author}
  {\bibfnamefont {M.~I.}\ \bibnamefont {Katsnelson}}, \bibinfo {author}
  {\bibfnamefont {I.~V.}\ \bibnamefont {Grigorieva}}, \bibinfo {author}
  {\bibfnamefont {S.~V.}\ \bibnamefont {Dubonos}}, \ and\ \bibinfo {author}
  {\bibfnamefont {A.~A.}\ \bibnamefont {Firsov}},\ }\href@noop {} {\bibfield
  {journal} {\bibinfo  {journal} {Nature}\ }\textbf {\bibinfo {volume} {438}},\
  \bibinfo {pages} {197} (\bibinfo {year} {2005})}\BibitemShut {NoStop}%
\bibitem [{\citenamefont {Zhang}\ \emph {et~al.}(2005)\citenamefont {Zhang},
  \citenamefont {Tan}, \citenamefont {Stormer},\ and\ \citenamefont
  {Kim}}]{Graphe_Berryphase2}%
  \BibitemOpen
  \bibfield  {author} {\bibinfo {author} {\bibfnamefont {Y.~B.}\ \bibnamefont
  {Zhang}}, \bibinfo {author} {\bibfnamefont {Y.W.}\ \bibnamefont {Tan}},
  \bibinfo {author} {\bibfnamefont {H.~L.}\ \bibnamefont {Stormer}}, \ and\
  \bibinfo {author} {\bibfnamefont {P.}~\bibnamefont {Kim}},\ }\href@noop {}
  {\bibfield  {journal} {\bibinfo  {journal} {Nature}\ }\textbf {\bibinfo
  {volume} {438}},\ \bibinfo {pages} {201} (\bibinfo {year}
  {2005})}\BibitemShut {NoStop}%
\bibitem [{\citenamefont {Weng}\ \emph {et~al.}(2015)\citenamefont {Weng},
  \citenamefont {Fang}, \citenamefont {Fang}, \citenamefont {Bernevig},\ and\
  \citenamefont {Dai}}]{WSM_PRX}%
  \BibitemOpen
  \bibfield  {author} {\bibinfo {author} {\bibfnamefont {H.}~\bibnamefont
  {Weng}}, \bibinfo {author} {\bibfnamefont {C.}~\bibnamefont {Fang}}, \bibinfo
  {author} {\bibfnamefont {Z.}~\bibnamefont {Fang}}, \bibinfo {author}
  {\bibfnamefont {B.~A.}\ \bibnamefont {Bernevig}}, \ and\ \bibinfo {author}
  {\bibfnamefont {X.}~\bibnamefont {Dai}},\ }\href@noop {} {\bibfield
  {journal} {\bibinfo  {journal} {Phys. Rev. X}\ }\textbf {\bibinfo {volume}
  {5}},\ \bibinfo {pages} {011029} (\bibinfo {year} {2015})}\BibitemShut
  {NoStop}%
\bibitem [{\citenamefont {Huang}\ and\ \citenamefont
  {et~al.}(2015)}]{TaAs_arXiv_NMR}%
  \BibitemOpen
  \bibfield  {author} {\bibinfo {author} {\bibfnamefont {X.}~\bibnamefont
  {Huang}}\ and\ \bibinfo {author} {\bibnamefont {et~al.}},\ }\href@noop {}
  {\bibfield  {journal} {\bibinfo  {journal} {arXiv:1503.01304}\ } (\bibinfo
  {year} {2015})}\BibitemShut {NoStop}%
\bibitem [{\citenamefont {Zhang}\ \emph {et~al.}(2015)\citenamefont {Zhang},
  \citenamefont {Yuan}, \citenamefont {Xu}, \citenamefont {Lin}, \citenamefont
  {Tong}, \citenamefont {Hasan}, \citenamefont {Wang}, \citenamefont {Zhang},\
  and\ \citenamefont {Jia}}]{TaAs_arXiv_Jia}%
  \BibitemOpen
  \bibfield  {author} {\bibinfo {author} {\bibfnamefont {C.}~\bibnamefont
  {Zhang}}, \bibinfo {author} {\bibfnamefont {Z.}~\bibnamefont {Yuan}},
  \bibinfo {author} {\bibfnamefont {S.}~\bibnamefont {Xu}}, \bibinfo {author}
  {\bibfnamefont {Z.}~\bibnamefont {Lin}}, \bibinfo {author} {\bibfnamefont
  {B.}~\bibnamefont {Tong}}, \bibinfo {author} {\bibfnamefont {M.~Z.}\
  \bibnamefont {Hasan}}, \bibinfo {author} {\bibfnamefont {J.}~\bibnamefont
  {Wang}}, \bibinfo {author} {\bibfnamefont {C.}~\bibnamefont {Zhang}}, \ and\
  \bibinfo {author} {\bibfnamefont {S.}~\bibnamefont {Jia}},\ }\href@noop {}
  {\bibfield  {journal} {\bibinfo  {journal} {arXiv:1502.00251}\ } (\bibinfo
  {year} {2015})}\BibitemShut {NoStop}%
\bibitem [{\citenamefont {Lv}\ and\ \citenamefont
  {et~al.}(2015{\natexlab{a}})}]{TaAs_ARPES1}%
  \BibitemOpen
  \bibfield  {author} {\bibinfo {author} {\bibfnamefont {B.~Q.}\ \bibnamefont
  {Lv}}\ and\ \bibinfo {author} {\bibnamefont {et~al.}},\ }\href@noop {}
  {\bibfield  {journal} {\bibinfo  {journal} {arXiv:1502.04684}\ } (\bibinfo
  {year} {2015}{\natexlab{a}})}\BibitemShut {NoStop}%
\bibitem [{\citenamefont {Lv}\ and\ \citenamefont
  {et~al.}(2015{\natexlab{b}})}]{TaAs_ARPES2}%
  \BibitemOpen
  \bibfield  {author} {\bibinfo {author} {\bibfnamefont {B.~Q.}\ \bibnamefont
  {Lv}}\ and\ \bibinfo {author} {\bibnamefont {et~al.}},\ }\href@noop {}
  {\bibfield  {journal} {\bibinfo  {journal} {arXiv:1503.09188}\ } (\bibinfo
  {year} {2015}{\natexlab{b}})}\BibitemShut {NoStop}%
\bibitem [{\citenamefont {Xu}\ and\ \citenamefont {et~al.}(2015)}]{NbAs_ARPES}%
  \BibitemOpen
  \bibfield  {author} {\bibinfo {author} {\bibfnamefont {S.}~\bibnamefont
  {Xu}}\ and\ \bibinfo {author} {\bibnamefont {et~al.}},\ }\href@noop {}
  {\bibfield  {journal} {\bibinfo  {journal} {arXiv:1504.01350}\ } (\bibinfo
  {year} {2015})}\BibitemShut {NoStop}%
\bibitem [{\citenamefont {Wang}\ \emph {et~al.}(2015)\citenamefont {Wang},
  \citenamefont {Zheng}, \citenamefont {Shen}, \citenamefont {Zhou},
  \citenamefont {Yang}, \citenamefont {Li}, \citenamefont {Feng},\ and\
  \citenamefont {Xu}}]{NbP_arXivWZ}%
  \BibitemOpen
  \bibfield  {author} {\bibinfo {author} {\bibfnamefont {Z.}~\bibnamefont
  {Wang}}, \bibinfo {author} {\bibfnamefont {Y.}~\bibnamefont {Zheng}},
  \bibinfo {author} {\bibfnamefont {Z.~X.}\ \bibnamefont {Shen}}, \bibinfo
  {author} {\bibfnamefont {Y.}~\bibnamefont {Zhou}}, \bibinfo {author}
  {\bibfnamefont {X.~J.}\ \bibnamefont {Yang}}, \bibinfo {author}
  {\bibfnamefont {Y.~P.}\ \bibnamefont {Li}}, \bibinfo {author} {\bibfnamefont
  {C.~M.}\ \bibnamefont {Feng}}, \ and\ \bibinfo {author} {\bibfnamefont
  {Z.~A.}\ \bibnamefont {Xu}},\ }\href@noop {} {\bibfield  {journal} {\bibinfo
  {journal} {arXiv:1506.00924}\ } (\bibinfo {year} {2015})}\BibitemShut
  {NoStop}%
\bibitem [{\citenamefont {Wan}\ \emph {et~al.}(2011)\citenamefont {Wan},
  \citenamefont {Turner}, \citenamefont {Vishwanath},\ and\ \citenamefont
  {Savrasov}}]{pair_PRB}%
  \BibitemOpen
  \bibfield  {author} {\bibinfo {author} {\bibfnamefont {X.~G.}\ \bibnamefont
  {Wan}}, \bibinfo {author} {\bibfnamefont {A.~M.}\ \bibnamefont {Turner}},
  \bibinfo {author} {\bibfnamefont {A.}~\bibnamefont {Vishwanath}}, \ and\
  \bibinfo {author} {\bibfnamefont {S.~Y.}\ \bibnamefont {Savrasov}},\
  }\href@noop {} {\bibfield  {journal} {\bibinfo  {journal} {Phys. Rev. B}\
  }\textbf {\bibinfo {volume} {83}},\ \bibinfo {pages} {205101} (\bibinfo
  {year} {2011})}\BibitemShut {NoStop}%
\bibitem [{\citenamefont {Nielsen}\ and\ \citenamefont
  {Ninomiya}(1983)}]{chiral_anomaly_PLB}%
  \BibitemOpen
  \bibfield  {author} {\bibinfo {author} {\bibfnamefont {H.~B.}\ \bibnamefont
  {Nielsen}}\ and\ \bibinfo {author} {\bibfnamefont {M.}~\bibnamefont
  {Ninomiya}},\ }\href@noop {} {\bibfield  {journal} {\bibinfo  {journal}
  {Phys. Lett. B}\ }\textbf {\bibinfo {volume} {130}},\ \bibinfo {pages} {389}
  (\bibinfo {year} {1983})}\BibitemShut {NoStop}%
\bibitem [{\citenamefont {Potter}\ \emph {et~al.}(2014)\citenamefont {Potter},
  \citenamefont {Kimchi},\ and\ \citenamefont {Vishwanath}}]{nov_p_nc}%
  \BibitemOpen
  \bibfield  {author} {\bibinfo {author} {\bibfnamefont {A.~C.}\ \bibnamefont
  {Potter}}, \bibinfo {author} {\bibfnamefont {I.}~\bibnamefont {Kimchi}}, \
  and\ \bibinfo {author} {\bibfnamefont {A.}~\bibnamefont {Vishwanath}},\
  }\href@noop {} {\bibfield  {journal} {\bibinfo  {journal} {Nat. Commun.}\
  }\textbf {\bibinfo {volume} {5}},\ \bibinfo {pages} {5161} (\bibinfo {year}
  {2014})}\BibitemShut {NoStop}%
\bibitem [{\citenamefont {Lundgren}\ \emph {et~al.}(2015)\citenamefont
  {Lundgren}, \citenamefont {Laurell},\ and\ \citenamefont
  {Fiete}}]{nov_p_PRB}%
  \BibitemOpen
  \bibfield  {author} {\bibinfo {author} {\bibfnamefont {R.}~\bibnamefont
  {Lundgren}}, \bibinfo {author} {\bibfnamefont {P.}~\bibnamefont {Laurell}}, \
  and\ \bibinfo {author} {\bibfnamefont {G.~A.}\ \bibnamefont {Fiete}},\
  }\href@noop {} {\bibfield  {journal} {\bibinfo  {journal} {Phys. Rev. B}\
  }\textbf {\bibinfo {volume} {90}},\ \bibinfo {pages} {165115} (\bibinfo
  {year} {2015})}\BibitemShut {NoStop}%
\bibitem [{\citenamefont {Ojanen}(2013)}]{nov_p_PRB2}%
  \BibitemOpen
  \bibfield  {author} {\bibinfo {author} {\bibfnamefont {T.}~\bibnamefont
  {Ojanen}},\ }\href@noop {} {\bibfield  {journal} {\bibinfo  {journal} {Phys.
  Rev. B}\ }\textbf {\bibinfo {volume} {87}},\ \bibinfo {pages} {245112}
  (\bibinfo {year} {2013})}\BibitemShut {NoStop}%
\bibitem [{\citenamefont {Ashby}\ and\ \citenamefont
  {Carbotte}(2013)}]{nov_p_PRB3}%
  \BibitemOpen
  \bibfield  {author} {\bibinfo {author} {\bibfnamefont {P.~E.}\ \bibnamefont
  {Ashby}}\ and\ \bibinfo {author} {\bibfnamefont {J.~P.}\ \bibnamefont
  {Carbotte}},\ }\href@noop {} {\bibfield  {journal} {\bibinfo  {journal}
  {Phys. Rev. B}\ }\textbf {\bibinfo {volume} {87}},\ \bibinfo {pages} {245131}
  (\bibinfo {year} {2013})}\BibitemShut {NoStop}%
\bibitem [{\citenamefont {Parameswaran}\ \emph {et~al.}(2015)\citenamefont
  {Parameswaran}, \citenamefont {Grover}, \citenamefont {Abanin}, \citenamefont
  {Pesin},\ and\ \citenamefont {Vishwanath}}]{nov_p_PRX}%
  \BibitemOpen
  \bibfield  {author} {\bibinfo {author} {\bibfnamefont {S.~A.}\ \bibnamefont
  {Parameswaran}}, \bibinfo {author} {\bibfnamefont {T.}~\bibnamefont
  {Grover}}, \bibinfo {author} {\bibfnamefont {D.~A.}\ \bibnamefont {Abanin}},
  \bibinfo {author} {\bibfnamefont {D.~A.}\ \bibnamefont {Pesin}}, \ and\
  \bibinfo {author} {\bibfnamefont {A.}~\bibnamefont {Vishwanath}},\
  }\href@noop {} {\bibfield  {journal} {\bibinfo  {journal} {Phys. Rev. X}\
  }\textbf {\bibinfo {volume} {4}},\ \bibinfo {pages} {031035} (\bibinfo {year}
  {2015})}\BibitemShut {NoStop}%
\bibitem [{\citenamefont {Fukushima}\ \emph {et~al.}(2008)\citenamefont
  {Fukushima}, \citenamefont {Kharzeev},\ and\ \citenamefont
  {Warringa}}]{CME_PRD}%
  \BibitemOpen
  \bibfield  {author} {\bibinfo {author} {\bibfnamefont {K.}~\bibnamefont
  {Fukushima}}, \bibinfo {author} {\bibfnamefont {D.}~\bibnamefont {Kharzeev}},
  \ and\ \bibinfo {author} {\bibfnamefont {H.}~\bibnamefont {Warringa}},\
  }\href@noop {} {\bibfield  {journal} {\bibinfo  {journal} {Phys. Rev. D}\
  }\textbf {\bibinfo {volume} {78}},\ \bibinfo {pages} {074033} (\bibinfo
  {year} {2008})}\BibitemShut {NoStop}%
\bibitem [{\citenamefont {Kim}\ and\ \citenamefont {et~al.}(2013)}]{WAL_PRL}%
  \BibitemOpen
  \bibfield  {author} {\bibinfo {author} {\bibfnamefont {H.~J.}\ \bibnamefont
  {Kim}}\ and\ \bibinfo {author} {\bibnamefont {et~al.}},\ }\href@noop {}
  {\bibfield  {journal} {\bibinfo  {journal} {Phys. Rev. Lett.}\ }\textbf
  {\bibinfo {volume} {111}},\ \bibinfo {pages} {246603} (\bibinfo {year}
  {2013})}\BibitemShut {NoStop}%
\bibitem [{\citenamefont {Furuseth}\ and\ \citenamefont
  {Kjekshuh}(1964)}]{NbAs_crystal1}%
  \BibitemOpen
  \bibfield  {author} {\bibinfo {author} {\bibfnamefont {S.}~\bibnamefont
  {Furuseth}}\ and\ \bibinfo {author} {\bibfnamefont {A.}~\bibnamefont
  {Kjekshuh}},\ }\href@noop {} {\bibfield  {journal} {\bibinfo  {journal} {Acta
  Crystallogr}\ }\textbf {\bibinfo {volume} {17}},\ \bibinfo {pages} {1077}
  (\bibinfo {year} {1964})}\BibitemShut {NoStop}%
\bibitem [{\citenamefont {Boller}\ and\ \citenamefont
  {Parthe}(1963)}]{NbAs_crystal2}%
  \BibitemOpen
  \bibfield  {author} {\bibinfo {author} {\bibfnamefont {H.}~\bibnamefont
  {Boller}}\ and\ \bibinfo {author} {\bibfnamefont {E.}~\bibnamefont
  {Parthe}},\ }\href@noop {} {\bibfield  {journal} {\bibinfo  {journal} {Acta
  Crystallogr}\ }\textbf {\bibinfo {volume} {16}},\ \bibinfo {pages} {1095}
  (\bibinfo {year} {1963})}\BibitemShut {NoStop}%
\bibitem [{\citenamefont {Shekhar}\ and\ \citenamefont
  {et~al.}(2015)}]{NbP_arXiv}%
  \BibitemOpen
  \bibfield  {author} {\bibinfo {author} {\bibfnamefont {C.}~\bibnamefont
  {Shekhar}}\ and\ \bibinfo {author} {\bibnamefont {et~al.}},\ }\href@noop {}
  {\bibfield  {journal} {\bibinfo  {journal} {arXiv:1502.04361}\ } (\bibinfo
  {year} {2015})}\BibitemShut {NoStop}%
\bibitem [{\citenamefont {Ghimire}\ and\ \citenamefont
  {et~al.}(2015)}]{NbAs_jpcm}%
  \BibitemOpen
  \bibfield  {author} {\bibinfo {author} {\bibfnamefont {N.~J.}\ \bibnamefont
  {Ghimire}}\ and\ \bibinfo {author} {\bibnamefont {et~al.}},\ }\href@noop {}
  {\bibfield  {journal} {\bibinfo  {journal} {J. Phys.: Condens. Matter}\
  }\textbf {\bibinfo {volume} {27}},\ \bibinfo {pages} {152201} (\bibinfo
  {year} {2015})}\BibitemShut {NoStop}%
\bibitem [{\citenamefont {T.~M.~Tritt}\ \emph {et~al.}(1999)\citenamefont
  {T.~M.~Tritt}, \citenamefont {Littleton}, \citenamefont {IV}, \citenamefont
  {Feger},\ and\ \citenamefont {Kolis}}]{Zr/HfTe5_anom_PRB}%
  \BibitemOpen
  \bibfield  {author} {\bibinfo {author} {\bibfnamefont {N.~D.~Lowhorn}\
  \bibnamefont {T.~M.~Tritt}}, \bibinfo {author} {\bibfnamefont {R.~T.}\
  \bibnamefont {Littleton}}, \bibinfo {author} {\bibfnamefont {A.~Pope}\
  \bibnamefont {IV}}, \bibinfo {author} {\bibfnamefont {C.~R.}\ \bibnamefont
  {Feger}}, \ and\ \bibinfo {author} {\bibfnamefont {J.~W.}\ \bibnamefont
  {Kolis}},\ }\href@noop {} {\bibfield  {journal} {\bibinfo  {journal} {Phys.
  Rev. B}\ }\textbf {\bibinfo {volume} {60}},\ \bibinfo {pages} {7816}
  (\bibinfo {year} {1999})}\BibitemShut {NoStop}%
\bibitem [{\citenamefont {Wang}\ \emph {et~al.}(2013)\citenamefont {Wang},
  \citenamefont {Weng}, \citenamefont {Q.~Wu},\ and\ \citenamefont
  {Fang}}]{Cd3As2_theory_LMR}%
  \BibitemOpen
  \bibfield  {author} {\bibinfo {author} {\bibfnamefont {Z.}~\bibnamefont
  {Wang}}, \bibinfo {author} {\bibfnamefont {H.}~\bibnamefont {Weng}}, \bibinfo
  {author} {\bibfnamefont {X.~Dai}\ \bibnamefont {Q.~Wu}}, \ and\ \bibinfo
  {author} {\bibfnamefont {Z.}~\bibnamefont {Fang}},\ }\href@noop {} {\bibfield
   {journal} {\bibinfo  {journal} {Phys. Rev. B}\ }\textbf {\bibinfo {volume}
  {88}},\ \bibinfo {pages} {125427} (\bibinfo {year} {2013})}\BibitemShut
  {NoStop}%
\bibitem [{\citenamefont {Murakawa}\ and\ \citenamefont
  {et~al.}(2013)}]{delta_science}%
  \BibitemOpen
  \bibfield  {author} {\bibinfo {author} {\bibfnamefont {H.}~\bibnamefont
  {Murakawa}}\ and\ \bibinfo {author} {\bibnamefont {et~al.}},\ }\href@noop {}
  {\bibfield  {journal} {\bibinfo  {journal} {Science}\ }\textbf {\bibinfo
  {volume} {342}},\ \bibinfo {pages} {1490} (\bibinfo {year}
  {2013})}\BibitemShut {NoStop}%
\end{thebibliography}%

\end{document}